\documentclass[12pt]{iopart}

\usepackage{tabularx}
\usepackage{graphicx}
\usepackage[dvipsnames]{xcolor}
\usepackage{xspace}
\usepackage{multirow}
\usepackage{url}
\usepackage{rotating}
\newcommand{\algoname}{SHAPES\xspace}

\begin{document}

\title[Adaptive spline glitch removal]{Glitch subtraction from gravitational wave data using adaptive spline fitting}

\author{Soumya D.~Mohanty, Mohammad A.~T.~Chowdhury}

\address{Department of Physics and Astronomy, University of Texas Rio Grande Valley, One West University Blvd., Brownsville, Texas 78520, USA}
\eads{\mailto{soumya.mohanty@utrgv.edu}}
\eads{mohammadabuthaher.chowdhury01@utrgv.edu}

\begin{abstract}
Transient signals of instrumental and environmental origins (``glitches") in gravitational wave data elevate the false alarm rate of searches for astrophysical signals and reduce their sensitivity. Glitches that directly overlap astrophysical signals hinder their detection and worsen parameter estimation errors. 
As the fraction of data occupied by detectable astrophysical signals will be higher in next generation detectors, such problematic overlaps could become more frequent. These adverse effects of glitches can be mitigated by estimating and subtracting them out from the data, but their unpredictable waveforms and large morphological diversity pose a challenge. 
Subtraction of glitches using data from auxiliary sensors as predictors works but
not for the majority of cases. 
Thus, there is a need for nonparametric glitch mitigation methods that do not require auxiliary data, work for a large variety of glitches, and have minimal effect on astrophysical signals in the case of overlaps. In order to cope with the high rate of glitches, it is also desirable that such methods be computationally fast. We show that adaptive spline fitting, in which the placement of free knots is optimized to estimate both smooth and non-smooth curves in noisy data, offers a promising approach to satisfying these requirements for  broadband short-duration glitches, the type that appear quite frequently. The method is demonstrated on  glitches drawn from three distinct classes in the Gravity Spy database as well as on the glitch that overlapped the binary neutron star signal GW170817. The impact of glitch subtraction on the GW170817 signal, or those like it injected into the data, is seen to be negligible.
\end{abstract}

\section{Introduction}
In a fairly short time since the first direct detection of a gravitational wave (GW) signal (GW150914) in 2015~\cite{PhysRevLett.116.061102} by the twin LIGO~\cite{advLIGO_CQG} detectors, GW astronomy has emerged as an information-rich field that will revolutionize our understanding of compact objects such as black holes and neutron stars. By now, the network of LIGO and Virgo~\cite{advVirgo_CQG} detectors has reported $90$ confirmed detections of GW signals from compact binary coalescences (CBCs) across the first observing run (O1)~\cite{abbott2019gwtc} to the third (O3)~\cite{abbott2021gwtc}. The majority of these are binary black hole (BBH) mergers but the haul also includes a binary neutron star (BNS) system (GW170817)~\cite{PhysRevLett.119.161101}.

The rate of detectable GW signals will grow as more detectors, namely KAGRA~\cite{KAGRA_Nature} and LIGO-India~\cite{doi:10.1142/S0218271813410101}, join the network and increase its distance reach for GW sources. Design studies are already underway for the successors to the current generation of GW detectors~\cite{0264-9381-27-19-194002,PhysRevD.91.082001,abbott2017exploring} with the goal of achieving an order of magnitude improvement in sensitivity across the current operational frequency band. In addition, next-generation detectors will seek to expand the operational range to lower frequencies ($\approx 1$~Hz), thereby increasing the duration of in-band GW signals across the board: for example, a BNS signal starting at $\approx 10$~Hz will last for days compared to the $\approx 1$~min for GW170817. Thus, future detectors will not only see a higher rate but also longer signals, raising the prospect~\cite{PhysRevD.86.122001} that there will be no data segment free of detectable GW signals. 

The false alarm rate  of searches for CBCs as well as generic short duration GW signals, or {\em bursts}, is dominated~\cite{Abbott_2018} by transient non-GW signals of instrumental or environmental origins, commonly called {\em glitches}. This is because glitches that populate the same frequency band as CBC or burst signals and happen to be transient in duration can falsely trigger the respective search pipelines. A glitch has a particularly adverse effect if it overlaps with a GW signal, as happened in the case of GW170817~\cite{PhysRevLett.119.161101}, and triggers the search pipeline to reject the glitch and possibly discard the signal. 
Even a non-overlapping  glitch can severely degrade parameter estimation if it is close enough to a GW signal~\cite{powell2018parameter}. In the third observing run of the LIGO and Virgo detectors, $\approx 20\%$ of detected GW signals overlapped with glitches~\cite{davis2022subtracting} due to the high glitch rate in Virgo. For future detectors, the frequency of accidental overlaps will be enhanced by the higher rate of detectable GW signals as well as, for CBC signals, their longer durations. 

Glitches have dissimilar and unpredictable waveforms but many of the observed ones tend to fall into distinct morphological classes.  This has motivated the investigation of automated glitch classification using machine learning where a range of different methods have been proposed, such as Support Vector Machine~\cite{Biswas2013}, t-Sne~\cite{BAHAADINI2018}, random forests~\cite{Biswas2013}, S-means~\cite{Mukherjee_2010}, and Deep Convolutional Neural Networks~\cite{Daniel2018}.
The Gravity Spy~\cite{zevin2017gravity} project uses a citizen science approach to engage the lay public in labeling glitches by visual inspection of their constant Q-transform (CQT)~\cite{brown1991calculation,0264-9381-21-20-024} time-frequency images. This has created a high quality training dataset for machine learning methods. By now,  more than $20$ named glitch classes are available in the Gravity Spy database, collected over multiple observing runs of the LIGO detectors~\cite{BAHAADINI2018}.

Several different approaches have been developed to mitigate the adverse effects of 
glitches on GW searches. 
GW search pipelines typically compute secondary functionals, called {\em vetoes}, of the data besides the primary detection statistic that help in distinguishing genuine GW signals from glitches. 
A well-known example is 
the Chi-square veto~\cite{PhysRevD.71.062001} used in CBC search pipelines. 
For LIGO-Virgo data, a set of Data Quality flags have been developed that use information from a large number of auxiliary sensors to quantify the safety of analyzing a given segment of GW strain data~\cite{abbott2016characterization}. 
For glitches that overlap a GW signal, the gating~\cite{gating_glitch} method removes the rectangular time-frequency block, or just the time interval, containing an identified glitch from the data. 
Cross-channel regression using data from auxiliary sensors ~\cite{1999gr.qc.....9083A,driggers2012active,0264-9381-32-16-165014,PhysRevD.101.042003} has been used to reduce excess broadband noise and a few types of glitches~\cite{offline_noise_subtract}. 

A relatively recent approach is that of estimating the waveform of a glitch from the data time series itself and subtracting it out. Glitch subtraction was of critical importance in the case of GW170817 and has been shown to be an important requirement in reducing bias in the estimation of GW signal parameters~\cite{Pankow_GW170817_Glitch_subtract}. 
The GW170817 glitch subtraction was carried out using the multi-detector BayesWave pipeline~\cite{cornish2015bayeswave,PhysRevD.103.044013}, which has also been used for other types of glitches~\cite{davis2022subtracting}. 
Another method, Glitschen~\cite{merritt2021transient}, follows the approach of constructing parametrized waveform models for identified glitch classes using principal component analysis of training sets. A strong motivation for developing  glitch estimation and subtraction methods is that one could, in principle, preprocess the data to clean out every sufficiently loud glitch of a known type. As exemplified by GW170817, where prior subtraction of the loud glitch would have kept the search pipeline from discarding the signal, this would make glitch rejection in all downstream GW searches safer.

In this paper, we present a method for the estimation and subtraction of broadband, short-duration glitches that have appeared frequently in the observation runs of the LIGO detectors.  The method is computationally cheap, works with single-detector data, does not require a training set of pre-identified glitches, and is not predicated on auxiliary sensor data. The core component of the method is \algoname (Swarm Heuristics based Adaptive and Penalized Estimation of Splines), an adaptive spline curve fitting algorithm introduced in~\cite{mohanty2020adaptive}\footnote{The \algoname code is available from the Github repository \texttt{mohanty-sd/SHAPES.git}.}. \algoname uses splines with free placement of knots to fit both smooth and non-smooth curves in noisy data. In particular, point discontinuities in the curve or its derivatives (up to some order) can be accommodated in the fit by allowing knots to merge. The ability to handle both sharp and slow changes in a curve is a built-in form of multiresolution analysis in SHAPES and a critical requirement for effective estimation of broadband glitches. 
 We examine the performance of our glitch subtraction method on the GW170817 glitch in LIGO-Livingston data and instances of glitches from three morphologically distinct classes, namely, {\em Blip}, {\em Koi Fish}, and {\em Tomte}, in 
 the Gravity Spy database.  In each of the latter three cases, we inject a BNS  signal overlapping with the glitch to mimic the case of GW170817. We find that the impact of glitch subtraction on the signals, real or injected, is negligible.

 The rest of the paper is organized as follows. Sec.~\ref{sec:shapes} reviews \algoname with the goal of  providing a self-contained description of the algorithm that is pertinent to this paper. Further details, such as the motivation and justification for certain features of the algorithm, can be found in~\cite{mohanty2020adaptive}. Sec.~\ref{sec:data} describes the dataset used in this paper and the details of how \algoname is used for glitch subtraction.  Sec.~\ref{sec:results} presents the results. Our conclusions and discussion of future work are presented in Sec.~\ref{sec:conclusions}.

\section{Adaptive spline fitting: the \algoname algorithm}
\label{sec:shapes}
\algoname is derived under the following models for the noisy data, $\overline{y}$, and the signal $\overline{s}(\theta)$.
\begin{eqnarray}
\overline{y} & = & \overline{s}(\theta) + \overline{\epsilon}\;,
\label{eq:regressionModel}
\end{eqnarray}
where $\overline{y}$, $\overline{s}$, and $\overline{\epsilon}$ are row vectors with $N$ elements, $y_i = y(t_i)$ and $s_i(\theta) = s(t_i;\theta)$, $i = 0,1,\ldots,N-1$, are samples taken at $t_i = i/f_s$ with $f_s$ being the sampling frequency, and $\theta$ denotes the set of signal parameters that need to be estimated from the data. The noise samples,
$\epsilon_i$, are drawn independently from the zero mean and unit variance normal (Gaussian) probability density function $N(0,1)$.  The white Gaussian noise assumption does not entail a loss of generalization since GW data can always be whitened using
the estimated noise power spectral density (PSD). 

\begin{figure}
 \centering
     \includegraphics[scale=0.5]{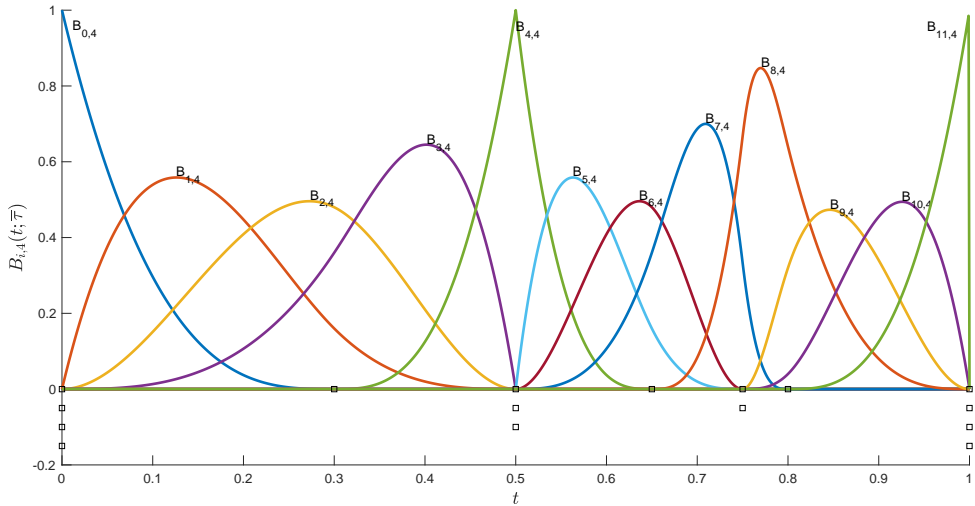}
     \caption{Cubic B-spline functions $B_{j,4}(t;\overline{\tau})$, $j = 0,1,\ldots,11$, for an 
     arbitrary choice of $16$ knots ($\overline{\tau}$) marked by squares.
     Knots with multiplicity $> 1$ result in B-splines that
     are discontinuous in value or derivatives.}
     \label{fig:bsplinefig}  
 \end{figure}

The signal $s(t;\theta)$ is assumed in \algoname to be a cubic spline, which is a polynomial of order $4$. The choice of a spline model as well as its order is an ad hoc one, with only an empirical justification, since a rigorous approach requires restricting the class of functions being estimated but this may be difficult for glitches. A cubic spline can be represented by a linear combination of B-spline functions~\cite{DEBOOR197250},
\begin{eqnarray}
s\left(t;\theta = \{\overline{\alpha},\overline{\tau}\}\right) &=& \sum_{j = 0}^{P-5}\alpha_j B_{j,4}(t;\overline{\tau})\;,
\label{eq:fittingFunction}
\end{eqnarray}
where $\overline{\alpha}= (\alpha_0,\alpha_1,\ldots,\alpha_{P-5})$, and  $\overline{\tau}= (\tau_0,\tau_1,\ldots,\tau_{P-1})$, $\tau_{i+1}\geq \tau_i$, is 
a sequence of $P$ {\em knots} that marks the end 
points of the contiguous intervals containing the 
 polynomial pieces of the spline.  Note that knots are allowed to be equal, leading to knots with multiplicity higher than one. (The knots $\tau_0$ and $\tau_{P-1}$ are repeated $3$ times each.) The number of B-splines in Eq.~\ref{eq:fittingFunction} corresponds to the number of independent parameters describing the spline after the continuity and differentiability constraints  on the polynomial pieces at the interior knots are taken into account.
 Repeating knots create discontinuity in either the value of a B-spline function or its derivatives (up to order $2$). This allows the $s(t;\theta)$ in Eq.~\ref{eq:fittingFunction} to model signals with point discontinuities in value or derivatives.  Fig.~\ref{fig:bsplinefig} illustrates cubic B-spline functions for an ad hoc knot sequence.
 
 The best fit spline parameters, $\widehat{\alpha}$ and $\widehat{\tau}$, are those that minimize a penalized least-squares function,
 \begin{eqnarray}
L_\lambda(\overline{\alpha},\overline{\tau}) & = & L(\overline{\alpha},\overline{\tau}) + \lambda R(\overline{\alpha})\;,
\label{eq:penalizedSpline}\\
L(\overline{\alpha},\overline{\tau}) & = & \sum_{i = 0}^{N-1} \left(y_i - s_i(\overline{\alpha},\overline{\tau})\right)^2\;,
\end{eqnarray}
where the penalty term,
\begin{eqnarray}
R(\overline{\alpha}) & = & \sum_{j=0}^{P-k-1} \alpha_j^2\;,
\label{eq:SHPS_penalty}
\end{eqnarray}
is found to be useful in the suppression of spurious clustering of the knots. These clusters are observed when the method tries to minimize  $L_\lambda(\overline{\alpha},\overline{\tau})$ by fitting out outlier data points arising from the noise alone. The strength of the penalty is controlled by the gain factor $\lambda$, with higher values of $\lambda$ leading to smoother estimates.
 
The optimization of $L_\lambda(\overline{\alpha},\overline{\tau})$ over the non-linear parameters $\overline{\tau}$
 has been a long-standing computational barrier~\cite{wold1974spline,burchard1974splines,jupp1978approximation,luo1997hybrid} for adaptive spline fitting. At the same time, 
the benefits of optimizing the placement of knots have also been demonstrated 
extensively~\cite{burchard1974splines,miyata2003adaptive}. It was shown in~\cite{galvez2011efficient}, and independently in~\cite{mohanty2012particle}, that Particle Swarm Optimization (PSO)~\cite{PSO,mohanty2018swarm},
a widely used nature-inspired metaheuristic for global optimization, 
has good performance on the free knot placement problem.  Moreover, being a continuous optimization method, PSO can
explore all arrangements of knots, including the ones where knots are sufficiently close to be merged into a single knot of higher multiplicity. This allows the 
fitting of functions that have a mix of smooth and non-smooth parts. 

There are many variations~\cite{engelbrecht2005fundamentals} among the algorithms that fall under the umbrella of the PSO metaheuristic but they all share the following common features. (i) They are continuous optimization methods that seek the global optimum of a function $f(\overline{x})$, $\overline{x}\in \mathbf{D}\subset \mathbf{R}^K$, called the {\em fitness function}, on an open subset $\mathbf{D}$, called the {\em search space}, of the space $\mathbf{R}^K$ of real $K$-tuples. (In the case of \algoname, the search space coordinates $\overline{x}$ are the knots $\overline{\tau}$.) The function is sampled at multiple locations, called {\em particles}, that move iteratively to explore the search space for the global optimum. The set of particles is called
a {\em swarm}.
(ii) The location of each particle is updated following a dynamical rule that typically uses the best location found by a particle in its history, called its {\em personal best}, and the best location found by the particles in its {\em neighborhood}, called its {\em local best}. Here, the fitness value at a location defines how good it is: for a minimization problem, the lower the fitness, the better the location. A common form of the dynamical rule computes the displacement of a particle by linearly combining three vectors: its previous displacement and the two vectors pointing from its current position to the personal and local best locations. However, the linear combination is performed with an independent random weight for each component of the latter two vectors.  (iii) Each particle explores the search space independently but is constantly attracted towards the personal and local bests. This leads to a form of communication between the particles that speeds up convergence to a promising region, followed by refinement of the solution until the iterations are terminated. 
 The PSO algorithm can avoid trapping by local minima due to both the randomness in the dynamical rule and the parallel exploration of the search space by the swarm.

The best location among all the particles at termination is the final solution found by the swarm for the global optimum. While there is no guarantee that the final solution is the true global optimum, the probability of successful convergence can be boosted exponentially by running multiple independent runs of PSO and picking the one with the best final solution. 
Most of the parameters involved in the PSO algorithm, such as the number of particles or the weights attached to the attractive forces, have very robust values across a wide variety of benchmark optimization problems~\cite{bratton2007defining} and rarely need to be changed. In our experience, there are typically only two quantities that need tuning: the number of iterations, $N_{\rm iter}$, to termination and the hyper-parameter $N_{\rm runs}$, the number of independent PSO runs. In this paper, we fix $N_{\rm iter}=2000$ and $N_{\rm runs}=8$ throughout. The number of particles is always set to $40$ and the settings for the remaining parameters, as well as the definition of the neighborhood used for the local best, are described in~\cite{mohanty2020adaptive}.

The description above was for the case where the number of knots, $P$, is 
fixed. The complete \algoname algorithm incorporates model selection using the Akaike Information Criterion (AIC)~\cite{akaike1998information}, where the optimum number of knots
minimizes,
\begin{eqnarray}
{\rm AIC} & = & 4P + L_\lambda(\widehat{\alpha},\widehat{\tau})\;.
\end{eqnarray}
 While, given sufficient computing resources, model selection could be performed over all values of $P$ until the minimum value of AIC is found, practical considerations dictate that the set of knot numbers used be a finite and small one. In this paper, for example, we use knot numbers in the set starting at $5$ and ending at $60$ in increments of $5$. It is important to note that this restriction of knot numbers is not a fundamental limitation but a technical one meant to manage the computational burden of model selection. Thus, the only significant free parameter that needs to be set by the user in the current version of \algoname is $\lambda$. 

Since \algoname assumes that the noise in the data is white, GW strain data 
must be whitened prior to glitch estimation and subtraction. The data conditioning steps involved are as follows (in sequential order). 
(a) Suppression of the seismic noise below $10$~Hz, (b) robust estimation of the  power spectral density (PSD) noise floor, (c) whitening of the noise floor using the estimated PSD~\cite{SMukhNonstat1}, and (d) automated identification of high-power narrowband noise  features (``lines") and their suppression using notch filters.  These steps are common to all GW search pipelines, so they do not need to be elaborated further here.

\section{Demonstration data}
\label{sec:data}
The glitches considered in this paper for demonstrating the performance of \algoname  are listed in Table~\ref{tab:Glitch Timeseries}. The corresponding GW strain data files can be located and downloaded from the Gravitational Wave Open Science Center (GWOSC)~\cite{vallisneri2015ligo} using the information provided in this table. We have used the standard $4096$~sec long GWOSC data files sampled at $4$~kHz. 

 The GW170817 glitch presents a particularly interesting example of the deleterious effect of glitches on GW searches. 
The GW signal appeared in both LIGO-Hanford (H1) and LIGO-Livingston (L1)
with a combined network signal to noise ratio (SNR) of $32.4$. Such a strong signal would have been detected easily in coincidence 
across L1 and H1 by the GW search pipelines in operation at the time. However, a coincident detection was prevented by a large overlapping glitch in L1 causing the release of only an unusual single-detector GW detection alert to the astronomical community. About $4.5$ hours elapsed between the initial alert and the release of the first skymap localizing GW170817 obtained by gating the glitch.

In addition to the GW170817 glitch, we have taken three representative glitches from the Blip, Koi Fish, and Tomte, classes in the Gravity Spy database~\cite{coughlin_scott_2021_5649212}. These glitches were selected by taking the loudest $5$ events, in terms of their signal-to-noise ratio (SNR) as given in the Gravity Spy database, for each class and then picking the first one in this list for which the corresponding GWOSC file had $100\%$ science data that was also reasonably stationary. As can be seen from Table~\ref{tab:Glitch Timeseries}, this results in the selected glitches spanning a wide range in SNR.
    \begin{table}[th]
        \centering
        \begin{tabular}{|c|c|c|c|c|}
              \hline
            Glitch Name & GPS start (sec) & SNR & Detector & run\\
            \hline
            GW170817 glitch  & 1187008880 & -- & L1 & O2 \\
            \hline
            Blip & 1182397347 & 109.1 & H1 & O2\\
            \hline
            Koi Fish & 1169847108 & 608.1 & H1 & O2 \\
            \hline
            Tomte & 1173086299 & 19.6 & H1 & O2\\
             \hline
        \end{tabular}
        \caption{Glitches considered in this paper along with their GPS start times, SNRs, the detectors in which they appeared, and the observation runs. For the Blip, Koi Fish, and Tomte glitches, the start times are taken from the Gravity Spy database. To the best of our knowledge, there is no SNR available in the literature for the GW170817 glitch.}
        \label{tab:Glitch Timeseries}
    \end{table}

After conditioning the data, we use the start time of a glitch, recorded in Table~\ref{tab:Glitch Timeseries}, to locate the glitch. Starting from the peak of the glitch, the data time series is scanned visually in both directions to identify a segment, containing the glitch, that tapers off  at both its boundaries to the general noise level of the conditioned data. 

To mimic the case of GW170817 and to study the effect of glitch subtraction on an overlapping GW signal, we injected a whitened restricted-2PN circularized binary inspiral signal with equal $1.4$~$M_\odot$ components in the conditioned data. The SNR  (in white noise with unit variance) of the injected signal is set at $37.3$, which is an ad hoc factor of $\sqrt{2}$ higher than the observed SNR of $26.4$ of GW170817 in L1~\cite{PhysRevLett.119.161101}.   The enhancement in SNR allows clearer visibility of the signal in time-frequency images while also posing a stronger challenge to \algoname in terms how well it ignores the GW signal when estimating a glitch. The segment  containing the glitch, taken from the conditioned data with the injected signal, is passed to \algoname for estimation of the glitch waveform followed by its subtraction.

\section{Results}
\label{sec:results}
In common with other papers on glitch estimation and subtraction, we present our results in the form of CQT time-frequency images and time series plots. These are obtained by taking projections of the data on a set of windowed sinusoids. The width of the window decreases with an increase in the carrier frequency, $f_c$, such that
$Q  = f_c/\Delta f$,
where $\Delta f$ is the $-3$~dB bandwidth of the Fourier transform of the window, remains constant. We use the CQT code provided in the 
 \texttt{librosa}~\cite{mcfee2015librosa} Python package for audio processing. For each glitch, we show CQTs of the conditioned data with injected signal  and the residual after subtraction of the glitch estimate.

\begin{figure}
    \centering
    \includegraphics[width=\linewidth]{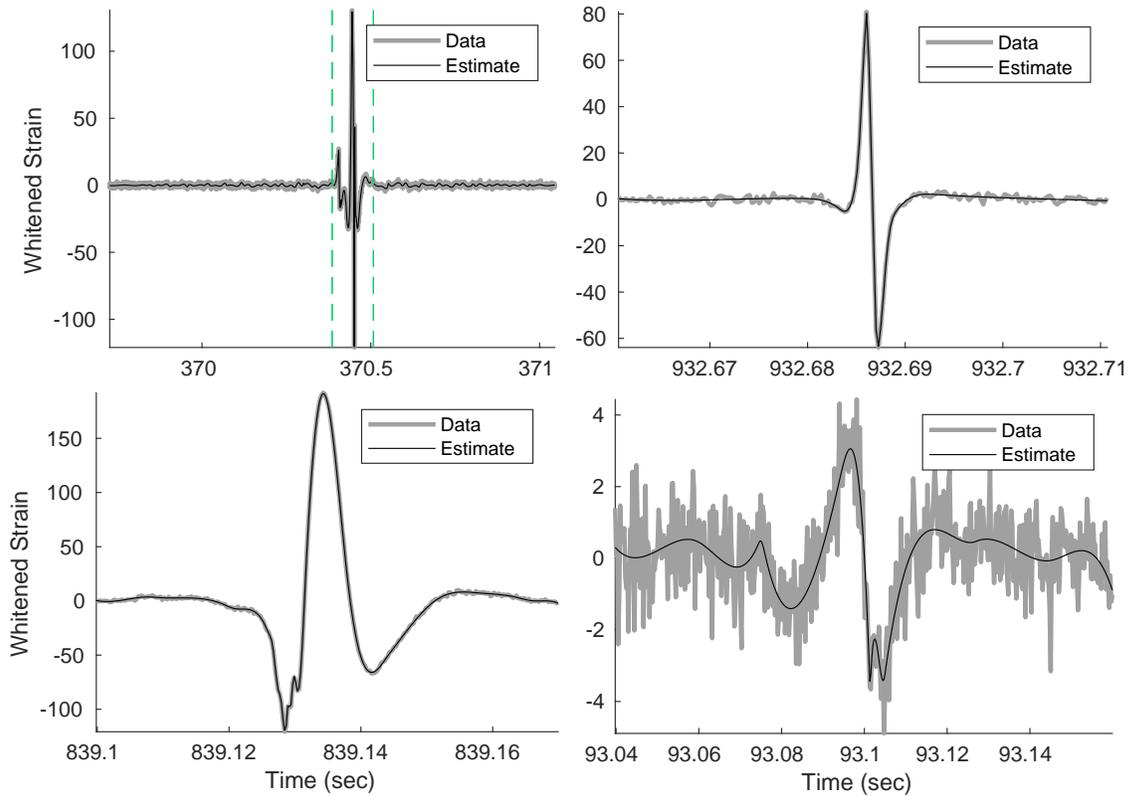}
    \caption{The conditioned strain data and the glitch waveform estimated by \algoname for each of the glitches considered in this paper. Top row: GW170817 (left) and
    Blip (right). Bottom row: Koi Fish (left) and Tomte (right). The X-axis in each plot shows the time (sec) since the start of the open data file containing the glitch as provided by GWOSC. For GW170817, the dashed vertical lines demarcate the three adjacent segments that were analyzed separately.}
    \label{fig:tserplot}
\end{figure}

Fig.~\ref{fig:tserplot} shows the data segments that were processed using 
\algoname and the corresponding estimated glitch waveforms. Except for GW170817, each segment was processed as a whole to obtain the glitch estimate. In the case of GW170817, \algoname was applied independently to three separate but contiguous time intervals to estimate the complete glitch. This was necessitated by the presence of extended wings, preceding and trailing the core broadband (and rapidly varying) part in the middle, that dominate the conditioned data for $\approx 0.5$~sec on each side. Applying \algoname to the complete segment would have required using a very large number of knots ($> 60$), making it unnecessarily expensive computationally given that splitting the segment achieves a good solution.

As mentioned in Sec.~\ref{sec:shapes}, the penalty gain $\lambda$ controls the smoothness of the estimate and is a user-specified parameter of the \algoname algorithm. Typically, when a glitch is loud and has a complex shape, $\lambda = 0.01$ allows \algoname to provide a better fit. For low SNR and simple glitch waveforms, or if the data is just plain white noise, $\lambda = 0.1$ does an adequate job. In general, estimates from \algoname are not sensitive to small variations of $\lambda$ around these values because the model selection is able to compensate for a lower value of $\lambda$ by selecting a higher knot number and vice versa. Without much fine tuning, we found that the values of $\lambda$ listed in Table~\ref{tab:settings} work well for the glitches studied in this paper. We have also listed in this table the number of knots for the best fit models selected by the AIC.
\begin{table}
    \centering
    \begin{tabular}{|c|c|c|}
    \hline
    Glitch Name & Penalty gain ($\lambda$) & Number of knots\\
    \hline
       GW170817 glitch  & 0.1, 0.01, 0.1 & 60,40,50 \\
       \hline
      Blip & 0.01 & 15\\
      \hline 
      Koi Fish & 0.01 & 30\\
      \hline
      Tomte & 0.1 & 15\\
      \hline
    \end{tabular}
    \caption{The penalty gain $\lambda$ used for the glitches and the number of knots in the best fit model. For the GW170817 glitch, there are three segments with the middle one containing the principal glitch and adjacent ones containing the wings. The penalty gains and best fit model are listed for all three segments in sequential order from left to right.}
    \label{tab:settings}
\end{table}

\begin{figure}[htb]
        \centering
         \includegraphics[width=0.8\linewidth]{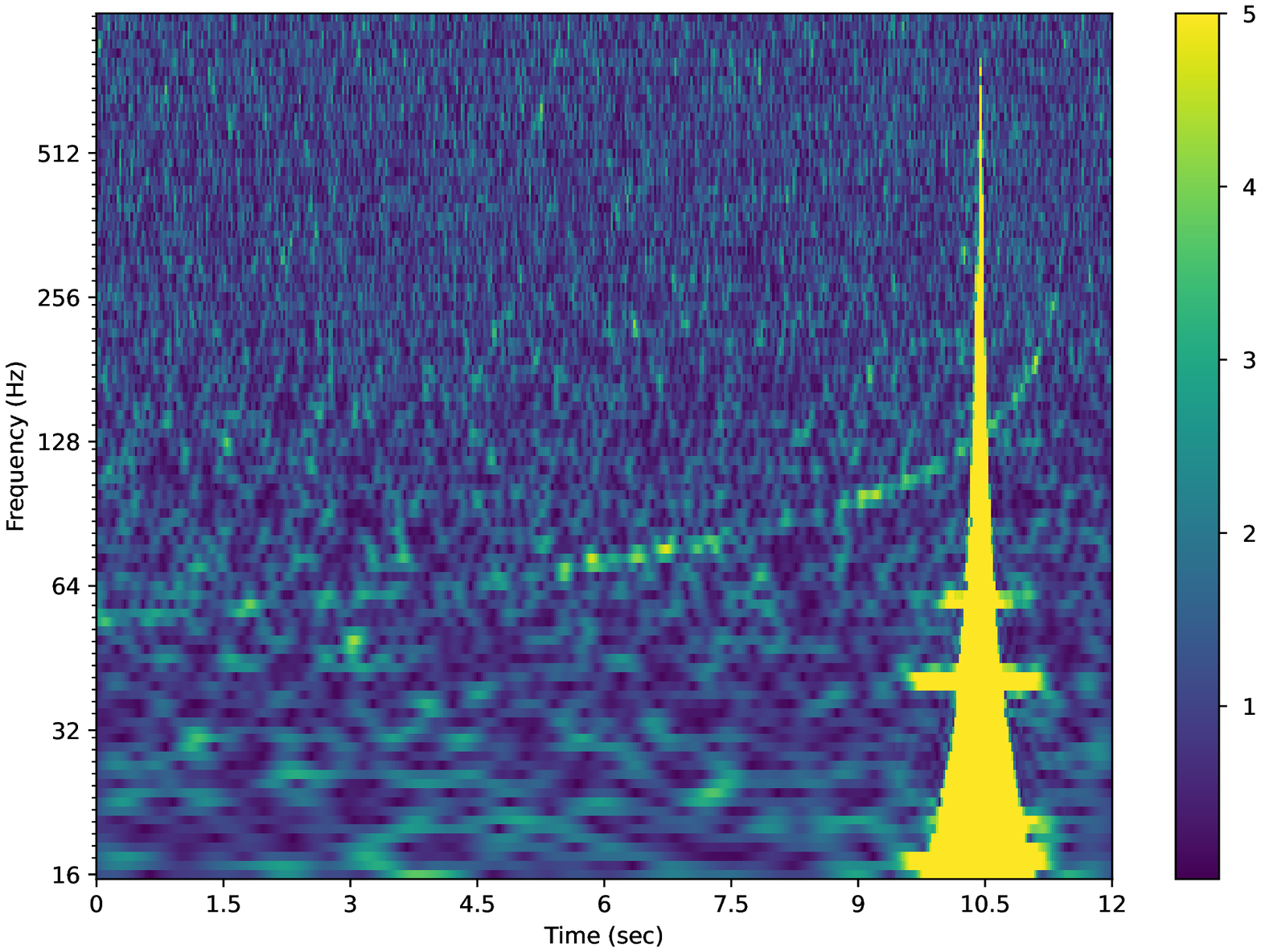}
         \includegraphics[width=0.8\linewidth]{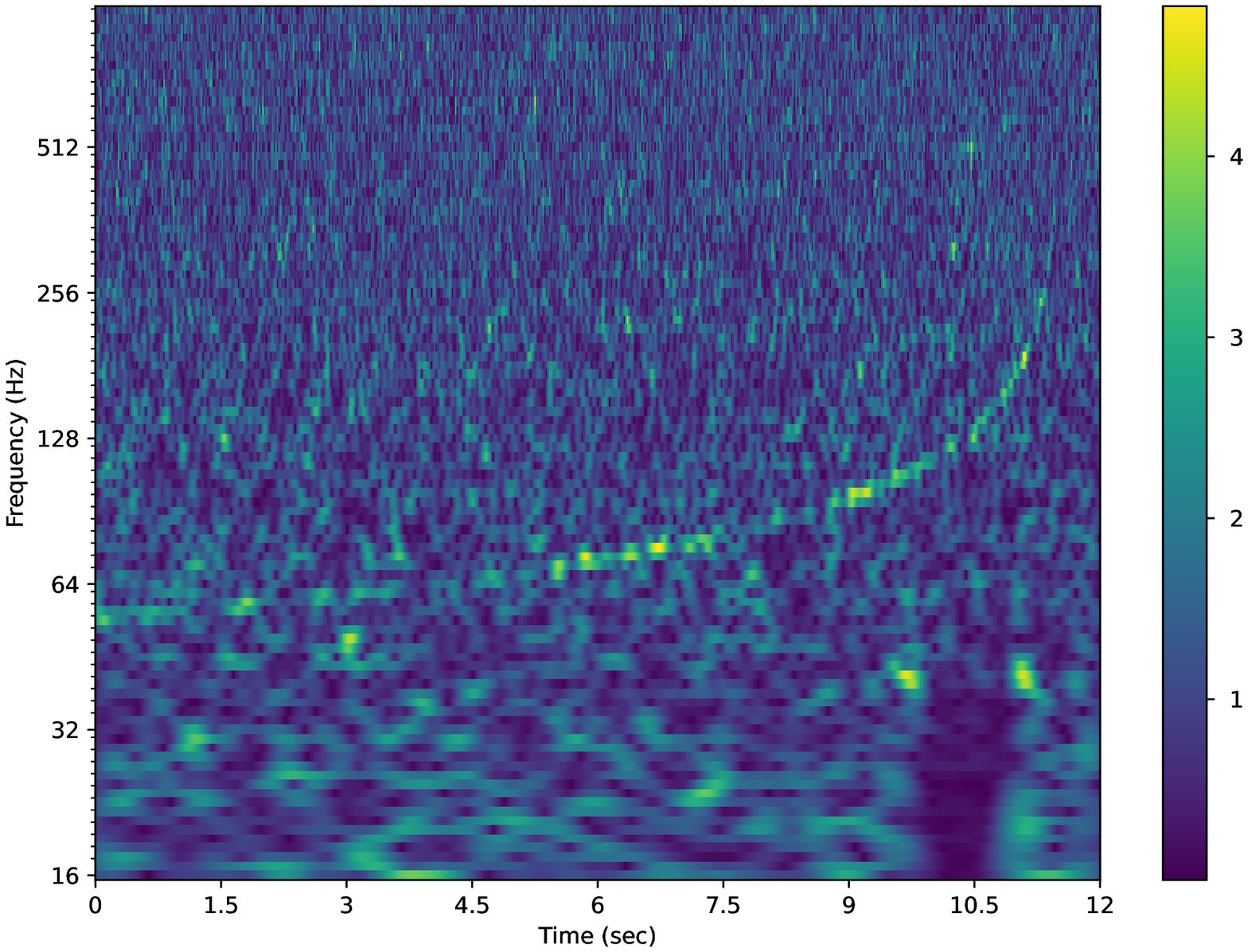}
        \caption{Subtraction of the GW170817 Glitch. The top and bottom panels show the CQT of the data and residual, respectively. The glitch is the vertical feature at $\approx 10.5$~sec. In order to show both the glitch and the signal in the same image, a threshold has been applied to the CQT as indicated by the maximum value in the colorbar of the top panel.}
        \label{fig:GW1701817}
  \end{figure}

\begin{figure}[htb]
        \centering
         \includegraphics[width=0.8\linewidth]{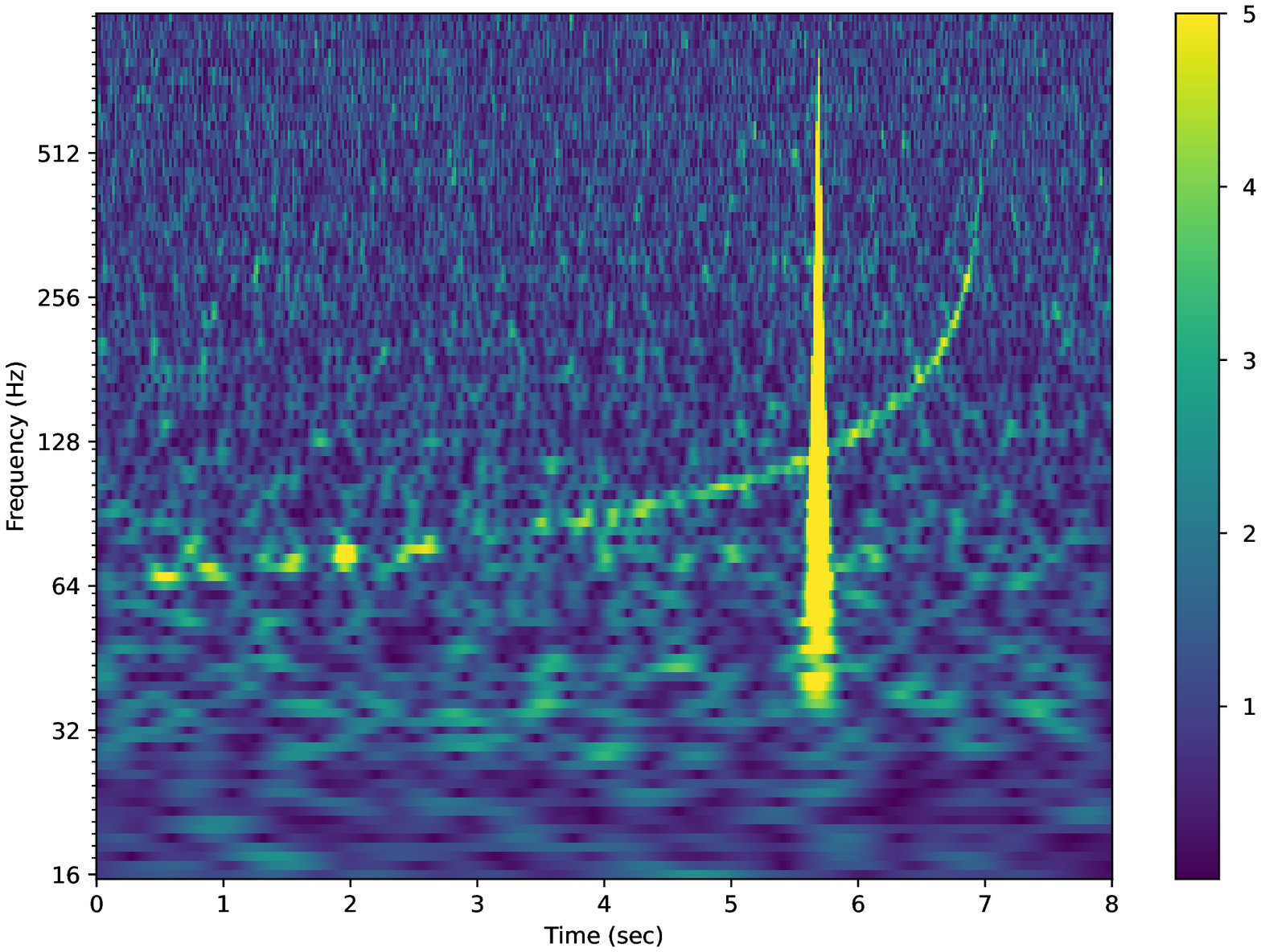}
         \includegraphics[width=0.8\linewidth]{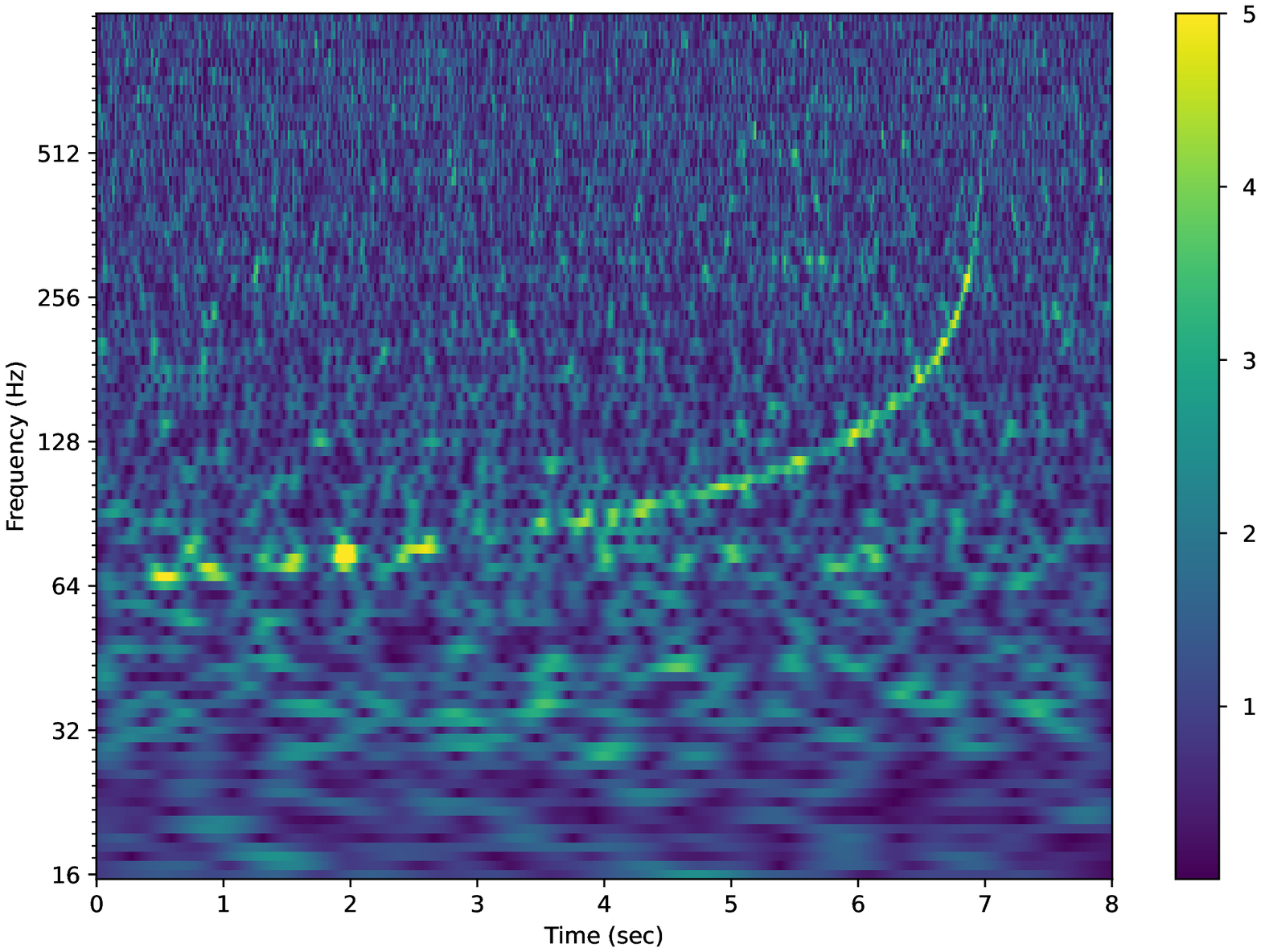}
        \caption{Subtraction of the Blip Glitch. The top and bottom panels show the CQT of the data and residual, respectively. The glitch is the vertical feature at $\approx 6$~sec. In order to show both the glitch and the signal in the same image, a threshold has been applied to the CQT as indicated by the maximum value in the colorbar of the top panel.}
        \label{fig:blip}
  \end{figure}

\begin{figure}[htb]
        \centering
         \includegraphics[width=0.8\linewidth]{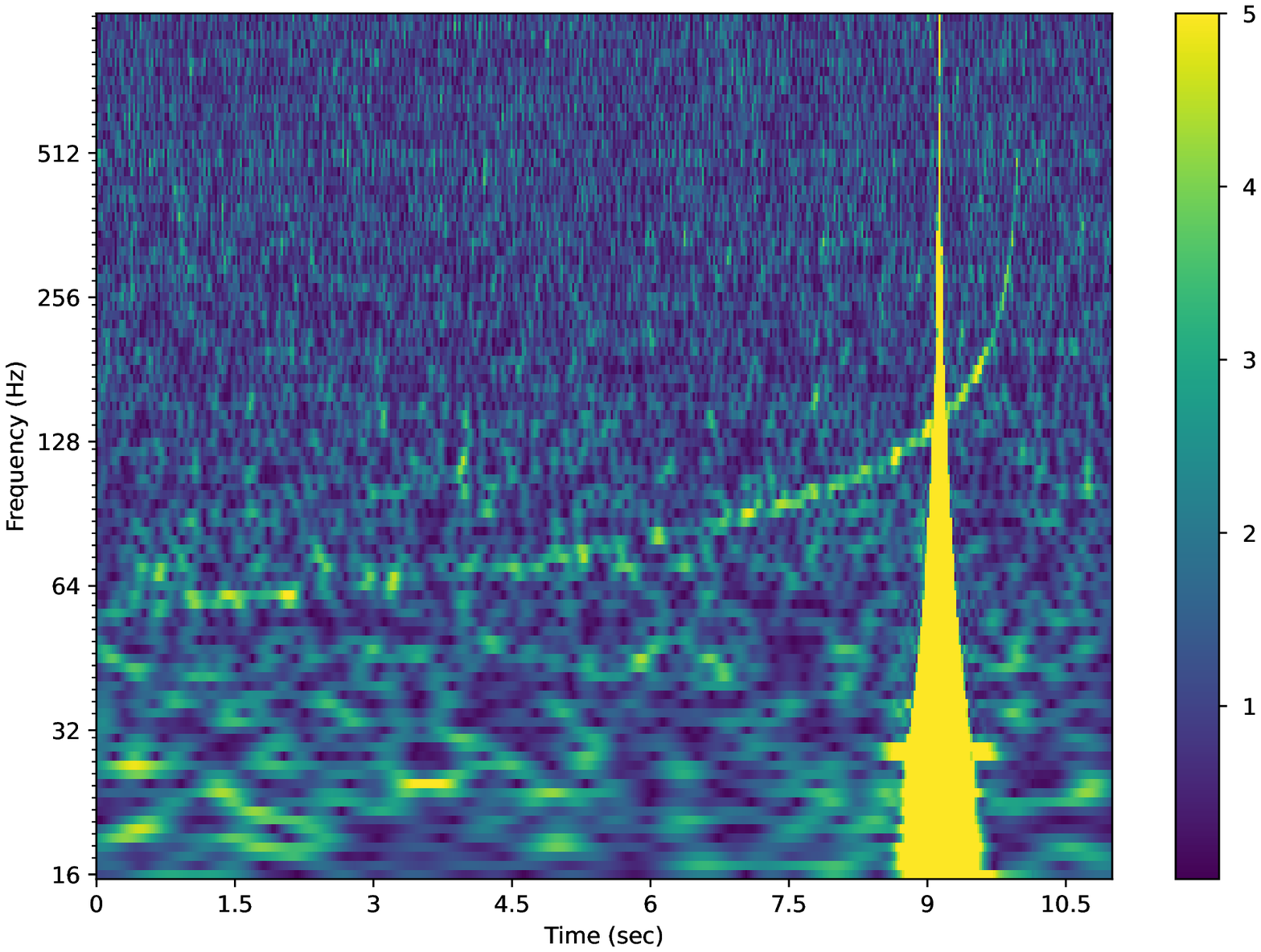}
         \includegraphics[width=0.8\linewidth]{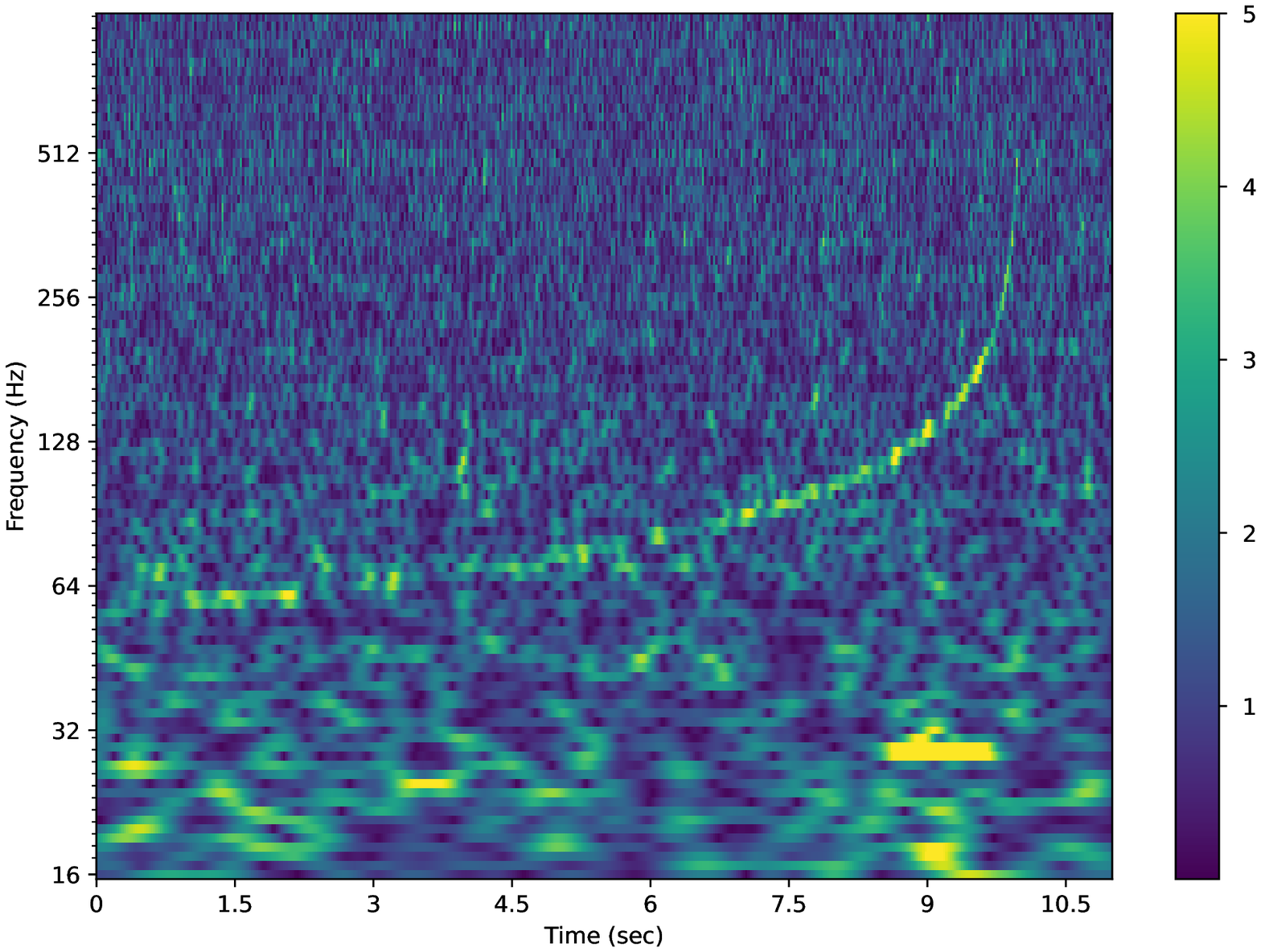}
        \caption{Subtraction of the Koi Fish glitch. The top and bottom panels show the CQT of the data and residual, respectively. The glitch is the vertical feature at $\approx 9.0$~sec. In order to show both the glitch and the signal in the same image, a threshold has been applied to the CQT as indicated by the maximum value in the colorbar of the top panel.}
        \label{fig:Koi_fish}
  \end{figure}

\begin{figure}[ht]
     \centering
     \includegraphics[width=.8\linewidth]{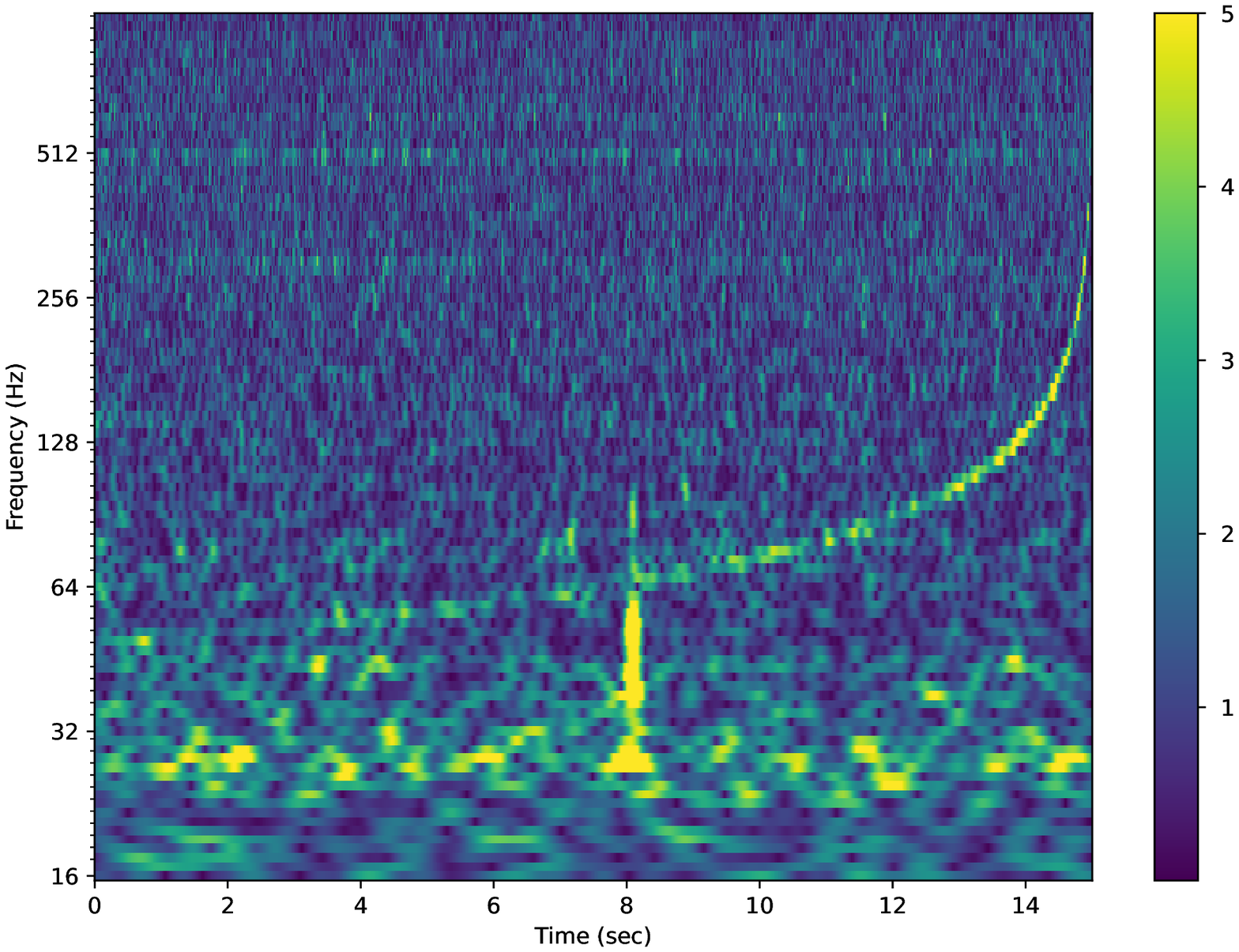} 
     \includegraphics[width=.8\linewidth]{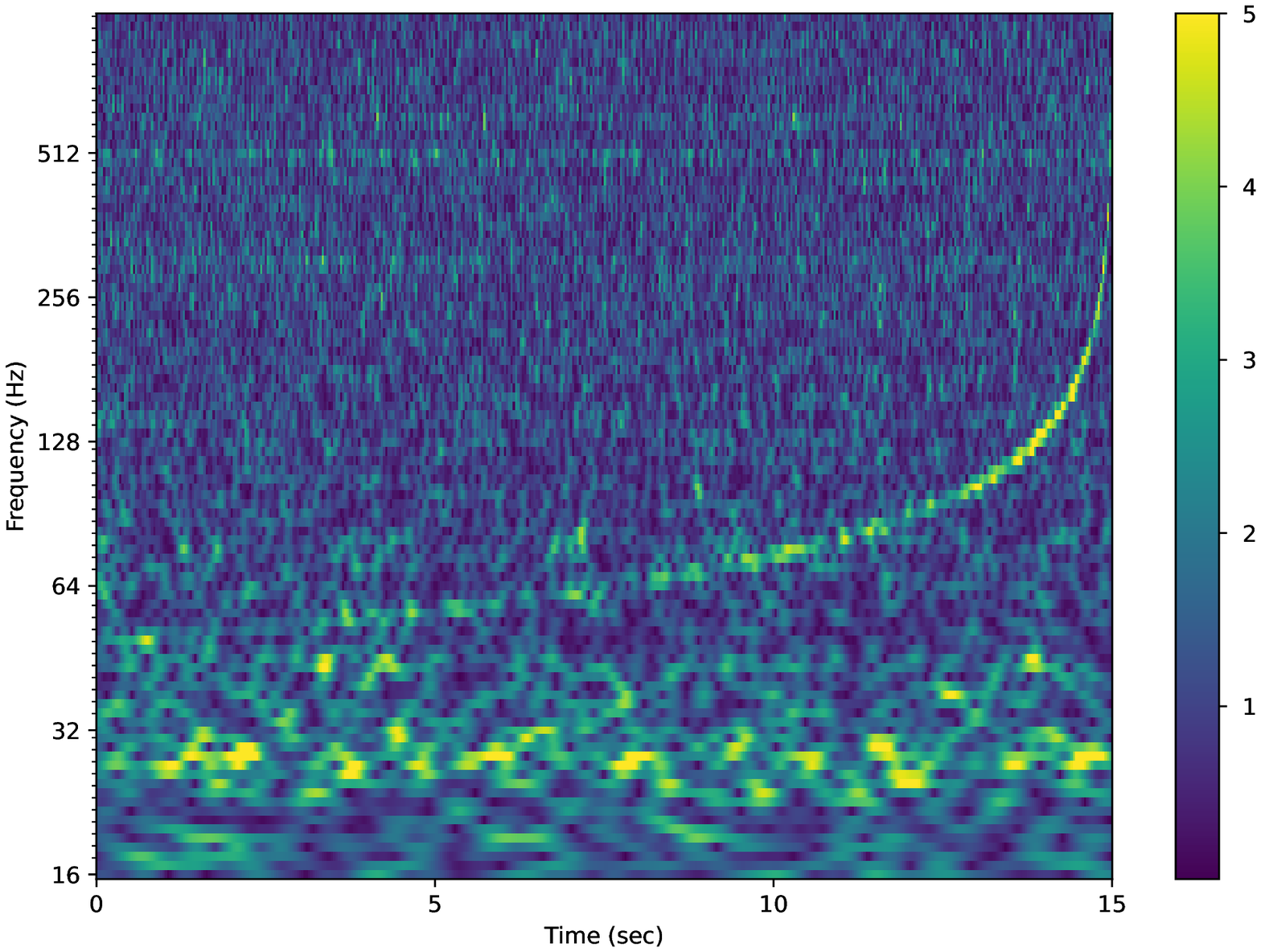} 
        \caption{Subtraction of the Tomte Glitch. The top and bottom panels show the CQT of the data and residual, respectively. The glitch is the vertical feature at $\approx 8.0$~sec. In order to show both the glitch and the signal in the same image, a threshold has been applied to the CQT as indicated by the maximum value in the colorbar of the top panel.}
        \label{fig:Tomte}
    \end{figure}

Fig.~\ref{fig:GW1701817} to Fig.~\ref{fig:Tomte} show the CQTs of the conditioned data and residuals after glitch subtraction for the glitches in the sequence GW170817, Blip, Koi Fish, and Tomte, respectively. In all cases, we see that the subtraction of the glitch does not affect the overlapping GW signal (real or injected) in any significant way. Some overfitting to the data, seen as very small CQT values, is visible in the residual for the GW170817 glitch at frequencies below $\approx 32$~Hz but this band has no overlap with the signal. The overfitted parts are the two wings of the GW170817 glitch mentioned earlier. The CQTs of the residuals for the Blip and Tomte glitches show near perfect removal of the glitch. (For Tomte, the coalescence time of the GW signal was kept further away from the glitch in order to create an overlap between the signal track and the glitch.) The residual for Koi Fish shows effective removal of the glitch with the exception of a transient and low frequency narrowband component. This leftover component does not overlap with the signal.
    
\begin{figure}
    \centering
    \includegraphics[width=.8\linewidth]{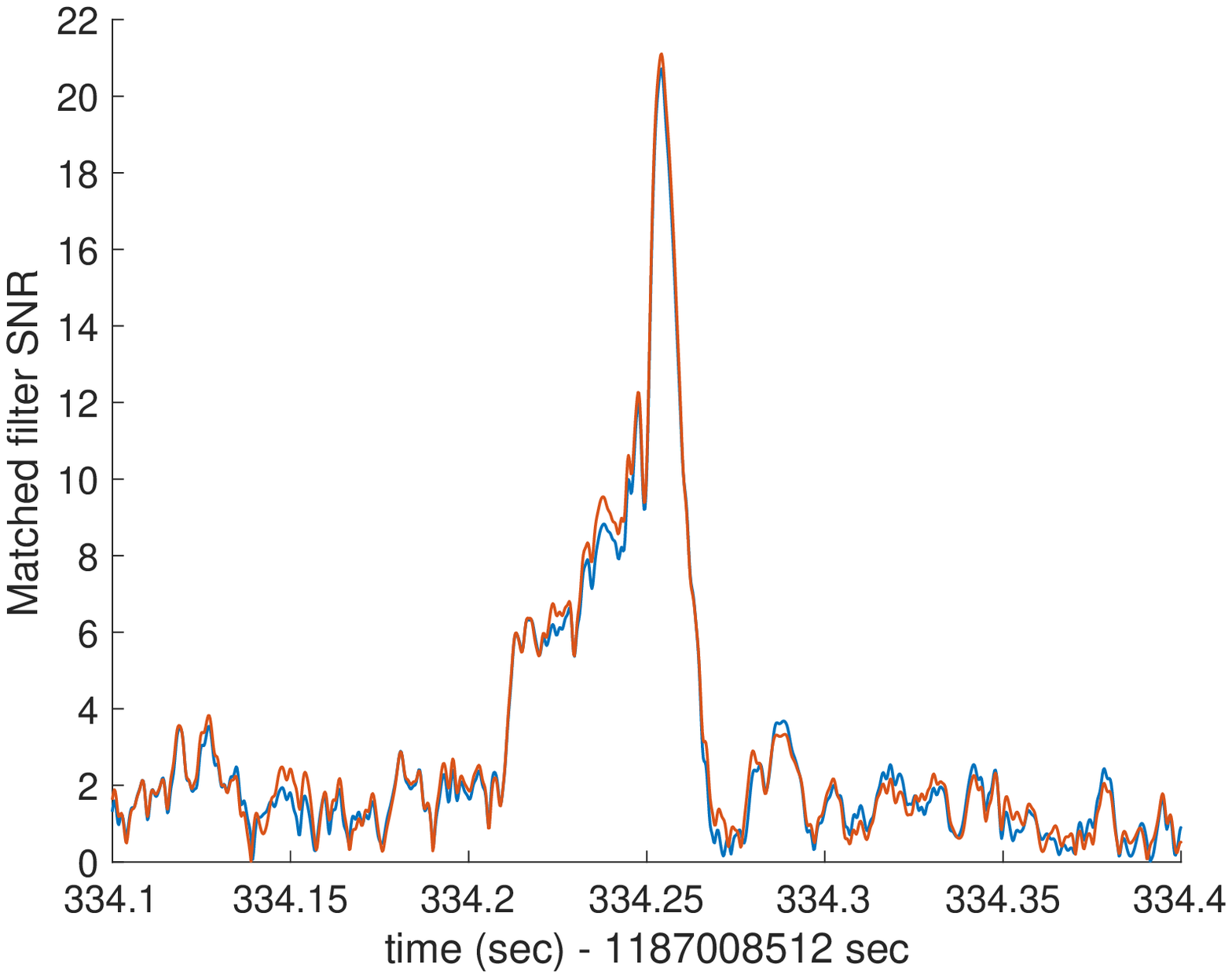}
    \caption{SNR time series outputs of matched filters with identical parameters on the  \algoname (blue) and BayesWave (red) residuals for the GW170817 L1 data. The peak values in both time series occur close to the time at which the instantaneous frequency of the signal crosses $35$~Hz.}
    \label{fig:comp_shapes_bwave}
\end{figure}

For the GW170817 glitch, it is possible to compare the performance of \algoname with BayesWave directly since the residual from the latter has been provided at GWOSC. (This is the version 2 data for this event from the L1 detector.) Fig.~\ref{fig:comp_shapes_bwave} shows the outputs of passing the two residuals through matched filters corresponding to the same BNS parameters. For this experiment, the matched filter templates are from the same family that we have used for injections but the parameters are tuned to be close to the ones estimated for GW170817, namely, the two masses are set at $1.46$~$M_\odot$ and $1.3$~$M_\odot$. (Since our conditioning pipeline is not identical to that used in LIGO, some tuning of the template parameter values is required to get a reasonable output SNR.) As can be seen, the peak values and their arrival times agree well with each other, demonstrating that \algoname has essentially the same effect on the signal as BayesWave.

The principal computational cost in \algoname is the global optimization of the fitness function in Eq.~\ref{eq:penalizedSpline}.  The time taken by the current \texttt{MATLAB}~\cite{matlab} code for a single PSO run on a segment with $\approx 300$ samples and knot numbers $P\in [10,60]$ (in steps of $5$) is 
$\approx 10$~min 
on an Intel Xeon E5 processor (clock rate $3$~GHz). The runtime increases with the number of knots used, mainly due to an increase in the number of B-spline functions that need to be computed. With a code currently under construction in the C language, and implementation of further hardware acceleration (e.g., using Graphics Processing Units), the runtime is expected to decrease substantially. We also note that the segments containing glitches can be processed in parallel since \algoname is a purely time-domain method. Hence, the computational cost will scale slower than linearly with the number of glitches when analyzing data containing multiple glitches.

\section{Discussion and Conclusions}
\label{sec:conclusions}
We have presented a new approach to glitch subtraction using an adaptive spline fitting method called \algoname. The method was demonstrated on the GW170817 glitch as well as other representative short duration and broadband glitches.  
In a single detector and in the absence of strong prior information about the signal, it is not possible to distinguish a GW signal from a glitch in the part where they overlap. Hence, it is expected that the signal power will be removed in that part along with the glitch when the latter is estimated and subtracted out. Nonetheless, as far as the BNS signal used in this paper is concerned, we observe very little impact on the signal across a wide range of glitch SNRs.   

\algoname is not well adapted to fitting highly oscillatory waveforms since these  are are not represented well by splines without using an inordinate number of knots. Therefore, the direct use of \algoname for glitches in the Gravity Spy database such as whistlers or wandering lines is not viable. However, chirp signals such as these could be estimated using the method proposed in~\cite{SEECR-PhysRevD.96.102008,Mohanty_eusipco_2018}, where adaptive splines figure indirectly in a non-linear signal model. This is an interesting direction that will be pursued in future work.

Other current limitations of \algoname, which are technical in nature, are that the penalty gain parameter $\lambda$ as well as the segment length to be processed must be specified by the user. The choice of the latter, along with the nature of the data, influences the number of knots used in the fit and led to the necessity of breaking up the data for the GW170817 glitch into three ad hoc parts. Work is in progress to address both of these limitations.

Our results show that \algoname is a promising addition to the toolbox of glitch subtraction methods that will become increasingly important as GW detectors become more sensitive. \algoname is computationally inexpensive, taking on the order of a few minutes for each glitch, and will be made much faster by planned code improvements. This could allow, in principle, the subtraction of a large number of broadband glitches of known types as part of data conditioning and provide significantly cleaner data for any type of GW search.

The data set used in this paper and codes for plotting glitch data and \algoname estimates are provided in a public data release on Zenodo~\cite{mohanty_soumya_d_2023_7592442}.

\ack{
S.D.M is supported by U.S. National Science Foundation (NSF) grant PHY-2207935 and partially supported by the U.S. Department of Defense grant W911NF2110169.
MATC acknowledges support from the Presidential Graduate Research Award at the University of Texas Rio Grande Valley. We acknowledge the Texas Advanced Computing Center (TACC) at the University of Texas at Austin (www.tacc.utexas.edu) for providing high performance computing resources.}

\end{document}